\documentclass{elsart5}

 \usepackage{graphicx}

\usepackage{amssymb}

\begin{document}

\begin{frontmatter}

\title{Magentic Ordering under Hydrostatic Pressure in Doped Spin Gap Systems ACu$_{1-x}$Mg$_x$Cl$_3$:\ A$=$Tl and K}

\author[aff1]{T. Ono}
\corauth[cor1]{}
\author[aff1]{H. Imamura}
\author[aff1]{J. Kawakami}
\author[aff1]{K. Goto}
\author[aff1]{H. Tanaka\corauthref{cor1}}
\ead{tanaka@lee.phys.titech.ac.jp}
\address[aff1]{Department of Physics, Tokyo Institute of Technology, Oh-okayama, Meguro-ku, Tokyo152-8551, Japan}
\received{12 June 2005}
\revised{13 June 2005}
\accepted{14 June 2005}


\begin{abstract}
Magnetic phase transitions under hydrostatic pressures in spin gap systems TlCu$_{0.988}$Mg$_{0.012}$Cl$_3$ and KCu$_{0.973}$Mg$_{0.027}$Cl$_3$ were investigated by magnetization measurements. 
The present doped systems exhibit impurity-induced magnetic orderings. With increasing pressure, ordering temperature $T_{\rm N}$ increases. With a further increase in pressure, the present systems undergo phase transitions to uniform antiferromagnetic phases due to the closing of the triplet gap in the intact dimers. The crossover from the impurity-induced ordered phase to the uniform antiferromagnetic phase occurs at $P\simeq 1.3$ kbar for TlCu$_{0.988}$Mg$_{0.012}$Cl$_3$.
\end{abstract}


\end{frontmatter}

Gapped ground states have been observed in many magnetic insulators. Most gapped ground states are composed of singlet spin dimers. Therefore, when a small number of nonmagnetic ions are substituted for magnetic ions, singlet spin dimers are partially broken, so that unpaired spins are produced near the nonmagnetic ions. Unpaired spins can interact through the effective exchange interaction $J_{\rm eff}$ that is mediated by the triplet excitations in intact dimers \cite{Sigrist,Mikeska}. This effective exchange interaction can cause the ordering of unpaired spins, which leads to the small staggered magnetic order in intact dimers. Such impurity-induced magnetic ordering is observed in many gapped spin systems doped with nonmagnetic impurities \cite{Masuda,Azuma,Uchiyama}. 

The present study is concerned with impurity- and pressure-induced magnetic orderings in ACu$_{1-x}$Mg$_x$Cl$_3$ with A$=$ Tl and K. The parent compounds TlCuCl$_3$ and KCuCl$_3$ are $S=1/2$ coupled spin dimer systems with antiferromagnetic intradimer exchange interactions $J/k_{\rm B} = 65.9$ K and 50 K, respectively \cite{Oosawa1,Cavadini}. The magnetic ground states are spin singlets with excitation gaps $\varDelta/k_{\rm B}$ = 7.5 K and 30.5 K \cite{Shiramura}. The gap decreases under hydrostatic pressure, and both systems undergo magnetic ordering \cite{Goto1,Goto2}. The critical pressures for TlCuCl$_3$ and KCuCl$_3$ are $P_{\rm c}=0.42$ kbar and 8.2 kbar, respectively. Because the effective interaction $J_{\rm eff}$ between unpaired spins is enhanced as the gap decreases \cite{Sigrist,Mikeska}, we expect an increase in $T_{\rm N}$ with pressure in ACu$_{1-x}$Mg$_x$Cl$_3$. To investigate the systematic change in the magnetic ordering in ACu$_{1-x}$Mg$_x$Cl$_3$, we carried out magnetization measurements under hydrostatic pressure.


Doped ACu$_{1-x}$Mg$_x$Cl$_3$ crystals were prepared by the vertical Bridgman method. The magnesium concentration $x$ was analyzed by ICP$-$OES. In the present study, we used samples with $x=0.012$ and 0.027 for A$=$Tl and K, respectively. 
The magnetizations were measured at temperatures down to 1.8 K under magnetic fields of up to 7 T using a SQUID magnetometer. Pressures of up to $\sim$9 kbar were applied using a cylindrical high-pressure cramp cell. A magnetic field was applied along the $[2,0,1]$ direction. As pressure-transmitting fluid, a mixture of Fluorinert FC70 and FC77, and Daphne oil 7373 were used. The pressure was calibrated with the superconducting transition temperature $T_{\rm c}$ of tin placed in the pressure cell. 
\begin{figure}[t]
\begin{center}
\includegraphics[scale =0.42]{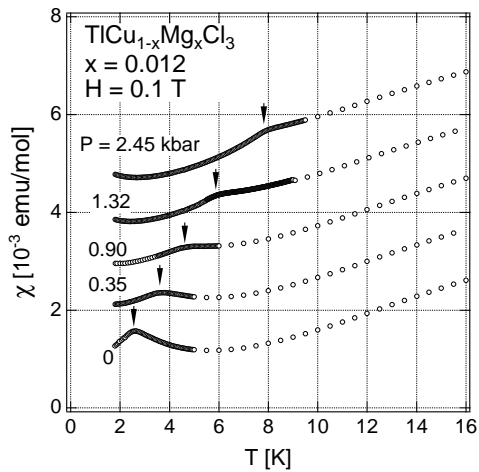}
\end{center}
\caption{Temperature dependences of magnetic susceptibilities in TlCu$_{0.988}$Mg$_{0.012}$Cl$_3$ measured at various pressures for $H\parallel [2,0,1]$. The values of the susceptibilities are shifted upward consecutively by $1\times 10^{-3}$ emu/mol with increasing pressure.}
  \label{fig:fig2.eps}
\end{figure}

\begin{figure}[t]
\begin{center}
\includegraphics[scale =0.44]{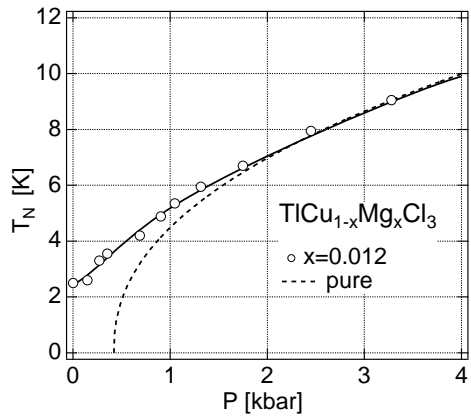}
\end{center}
\caption{Transition temperature $T_{\rm N}$ as function of pressure in TlCu$_{0.988}$Mg$_{0.012}$Cl$_3$. The solid line is a visual guide. The dashed line denotes the transition temperature of pressure-induced magnetic ordering in pure TlCuCl$_3$ \cite{Goto1}.}
  \label{fig:fig2.eps}
\end{figure}

\begin{figure}[t]
\begin{center}
\includegraphics[scale =0.43]{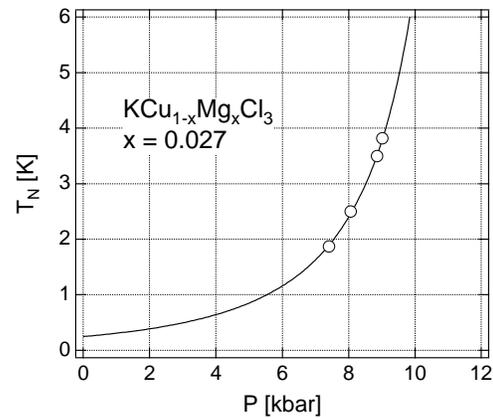}
\end{center}
\caption{Transition temperature $T_{\rm N}$ as function of pressure in KCu$_{0.973}$Mg$_{0.027}$Cl$_3$. The solid line is a visual guide.}
  \label{fig:fig2.eps}
\end{figure}


Figure 1 shows the temperature dependences of the magnetic susceptibilities in TlCu$_{0.988}$Mg$_{0.012}$Cl$_3$ measured at various pressures.
At ambient pressure, the magnetic susceptibility in TlCu$_{0.988}$Mg$_{0.012}$Cl$_3$ displays a broad maximum at $T\simeq 36$ K and then a decrease with decreasing temperature, as observed in pure TlCuCl$_3$. However, the magnetic susceptibility in the present doped system increases again below 5 K, and then exhibits a sharp cusplike anomaly at $T_{\rm N}=2.5$ K, which is indicative of magnetic ordering. The increase in $\chi$ below 5 K is attributed to the paramagnetic susceptibility of unpaired spins and their ordering gives rise to the decrease in $\chi$ below $T_{\rm N}=2.5$ K. The pressure dependence of $T_{\rm N}$ is plotted in Fig. 2, where we appended $T_{\rm N}$ of pressure-induced magnetic ordering in pure TlCuCl$_3$ \cite{Goto1}. The transition temperature $T_{\rm N}$ of impurity-induced magnetic ordering increases monotonically with pressure.  As mentioned above, the effective interaction $J_{\rm eff}$ between unpaired spins increases as the gap decreases \cite{Sigrist,Mikeska}. In the present doped system, the gap shrinks with increasing pressure. Thus, the $J_{\rm eff}$ increases with increasing pressure. This gives rise to the increase in $T_{\rm N}$ for impurity-induced magnetic ordering.

With increasing pressure, the height of the susceptibility cusp at $T_{\rm N}$ is suppressed and the cusplike anomaly changes to a bend anomaly at $P_0\simeq 1.3$ kbar. The bend anomaly at $T_{\rm N}$ is characteristic of pressure-induced magnetic transition to the uniform antiferromagnetic phase, as observed in pure TlCuCl$_3$ \cite{Goto1}. In the uniform antiferromagnetic phase, the triplet excitation gap in intact dimers closes completely. The cusplike anomaly in susceptibility is characteristic of impurity-induced magnetic phase transition. Since the changes in the susceptibility anomaly and $T_{\rm N}$ at $P_0$ are not rapid but gradual, we can deduce that the crossover between impurity- and pressure-induced uniformly ordered phases occurs at $P_0\simeq 1.3$ kbar. This crossover pressure is more than twice the critical pressure $P_{\rm c}=0.42$ kbar of the pressure-induced magnetic ordering in TlCuCl$_3$ \cite{Goto1}. The triplet gap remains even in the doped TlCu$_{1-x}$Mg$_x$Cl$_3$ system and increases with $x$ \cite{Oosawa1}. 
We infer that the increase in the triplet gap with doping is responsible for the increase in the crossover pressure $P_0$, against $P_{\rm c}=0.42$ kbar for TlCuCl$_3$.

For KCu$_{0.973}$Mg$_{0.027}$Cl$_3$, no impurity-induced magnetic ordering was observed down to 1.8 K at ambient pressure. This is because the effective interaction $J_{\rm eff}$ is much smaller than that in TlCu$_{0.988}$Mg$_{0.012}$Cl$_3$ due to larger triplet gap. However, with increasing pressure, ordering temperature $T_{\rm N}$ increases. For $P > 7.4$ kbar, $T_{\rm N}$ becomes higher than 1.8 K and increases rapidly with pressure, as shown in Fig. 3. This behavior is attributed to the exponential increase in $J_{\rm eff}$ due to the shrinkage of the triplet gap with pressure.

This work was supported by a Grant-in-Aid for Scientific Research and the 21st Century COE Program ``Nanometer-Scale Quantum Physics'' at Tokyo Tech., both from the Ministry of Education, Culture, Sports, Science and Technology of Japan.

\end{document}